\documentclass[twocolumn,showpacs,aps,prc,amsmath,amssymb]{revtex4-1}

\usepackage{graphicx}
\usepackage{dcolumn}
\usepackage{bm}
\begin{document}
\title{Complex fragment emission in low energy light-ion reactions}
\author{S. Kundu}
\email{skundu@vecc.gov.in, samir.kundu@gmail.com}
\author{C. Bhattacharya}%
\author{K. Banerjee}
\author{T. K. Rana}
\author{S. Bhattacharya}
\author{A. Dey}
\author{T. K. Ghosh}
\author{G. Mukherjee}
\author{J. K. Meena}
\author{P. Mali}
\altaffiliation[Present address: ]{Dept. of Physics, University of North Bengal, Silliguri - 734013, INDIA}
\author{S. Mukhopadhyay}
\author{D. Pandit}
\author{H. Pai}
\author{S. R. Banerjee}
\author{D. Gupta}
\altaffiliation[Present address: ]{Dept. of Physics and Centre for Astroparticle Physics and Space Science, Bose Institute, Bidhan Nagar, Kolkata - 700091, INDIA}
\affiliation{ Variable Energy Cyclotron Centre, 1/AF, Bidhan Nagar, Kolkata - 700064, INDIA }
\author{P. Banerjee}
\affiliation{Presidency University, Kolkata - 700073, INDIA}
\author{Suresh Kumar}
\author{A. Shrivastava}
\author{A. Chatterjee}
\author{K. Ramachandran}
\author{ K. Mahata}
\author{S. K. Pandit}
\author{S. Santra}
\affiliation{Nuclear Physics Division, Bhabha Atomic Research Centre, Mumbai - 400085, INDIA}
\begin{abstract}
Inclusive energy spectra of  the complex fragments (3 $\leq$ Z $\leq$ 5) emitted in the reactions $^{12}$C (77 MeV)+ $^{28}$Si, $^{11}$B (64 MeV)+ $^{28}$Si and $^{12}$C (73 MeV)+ $^{27}$Al (all having the same excitation energy of $ \sim$ 67 MeV),  have been measured in the angular range of  10$^\circ$ $\lesssim \theta_{lab} \lesssim$ 60$^\circ$. The fully energy damped  (fusion-fission)  and the partially energy damped  (deep inelastic) components  of the  fragment energy spectra   have been extracted. It has been found that the yields of  the fully energy damped fragments  for all the above reactions are in conformity with the respective statistical model predictions. The time scales of various deep inelastic fragment emissions have been extracted from the angular distribution data. The angular momentum dissipation in deep inelastic collisions has been estimated from the data and it has been found to be close to the corresponding sticking limit value.
\end{abstract}

\maketitle

\section{\label{sec1:intro}Introduction}
The phenomenon of complex fragment emission in low- and intermediate-energy nucleus-nucleus collisions has been a subject of intense theoretical \cite{more1,more2,cha,cha2,kalan1,Sanders91,dha2,Szanto97} and experimental \cite{sob1,sob2,mcm,koz,gro1,gro,kwi,adem,shro1,shro2,sob3,gro1,caval,carlin,Sanders99,samir1, ade1,ad2,szliner,bhat05,pop,bhat2002,bhat2004,bhat96,bec98,beck96,barrow,farrar96, ranjos93,cbeck93,shapi79,shapi,shapi82,shapi84,shiva,dunn} studies for the last few decades. It is nowadays known from these studies that the origin of the complex fragments may be broadly classified into two major categories, i.e., fusion-fission (FF) and non-fusion (deep inelastic, quasi elastic, breakup, etc.) processes. A large part of the above studies have been devoted to understand the mechanism of complex fragment  emission in fusion-fission process for both heavy (typically, A$_{projectile}$ + A$_{target} \ \gtrsim$ 60) as well as light compound systems \cite{more1,more2,cha,cha2,kalan1,sob1,sob2,mcm,koz,gro1,gro,kwi,adem, gro1, caval, carlin, Sanders99,samir1, ade1, ad2, szliner, bhat05,pop, bhat2002,bhat2004,bhat96,bec98, beck96,barrow,farrar96,ranjos93,cbeck93,shapi79,shapi,shapi82,shapi84,shiva,dunn}. On the other hand, the properties of deep inelastic (DI) reactions have been studied in details mostly for heavier systems in the past decades (see, for example, \cite{shro1,shro2,sob3} and references therein) to extract important information about the origins of nuclear relaxation processes, and the data on DI reactions for lighter systems are rather scarce \cite{ad2,pop,bhat2002,bhat2004}. This might be due to the fact that, unlike in the case of heavy systems, the distinction between the DI and the FF processes is rather difficult for light systems, as in the later case there is strong overlap in the elemental distributions of the fragments originating from the two processes.
The scenario becomes further complicated particularly for the reactions involving $\alpha$ - cluster nuclei, where nuclear structure is also  known to play an important role in the equilibrium emission of  complex fragments. In these cases, in addition to the standard fusion-fission route of fragment emission, the projectile and the target  have a finite probability to form a long-lived dinuclear composite, which directly undergoes scission (without the formation of the fully equilibrated compound nucleus) to emit complex fragments. This process, termed as nuclear orbiting \cite{shiva}, has been shown to contribute significantly to the fragment yield in many reactions involving light $\alpha$ - cluster nuclei (e.g., $^{16}$O + $^{12}$C \cite{samir1}, $^{20}$Ne + $^{12}$C \cite{ade1, bhat05, shapi79, shapi}, $^{24}$Mg + $^{12}$C \cite{dunn}, $^{28}$Si + $^{12}$C \cite{shapi82,shapi84} etc.).

In recent years, a few studies have been made on the $\alpha$-cluster  system   $^{40}$Ca$^* $ and the neighboring non-$\alpha$-cluster systems to look into the relationship between equilibrium emission of fragment (and \textit{vis-\`a-vis} orbiting) and $\alpha$-clustering.  From the study of fragment emission (6 $\le Z \le$ 8) in the inverse kinematical reaction $^{28}$Si +$^{12}$C  at energies 29.5 MeV $<$ E$_{c.m}$ $<$50 MeV \cite{shapi84}, it has been conjectured that orbiting played a crucial role in  fully energy-damped fragment emission. Even for the non~-~$\alpha$~-~cluster system with $A_{CN} \simeq 42$ ($^{28}$Si + $^{14}$N), where the number of open reaction channels was large  compared to that of $^{28}$Si + $^{12}$C \cite{beck94}, the  yields of fully energy-damped fragments (6 $\le Z \le$ 8) were found to have contributions, though smaller in magnitude, from  the orbiting process \cite{shiva2,shiva3,shiva4}.

 It will, therefore, be worthwhile to study the emission of lighter fragments ( $Z <$ 6) in particular, for systems around $A_{CN} \simeq 40$, to extract the contributions of different emission mechanisms, which will be partly complementary to the earlier measurements \cite{shapi84}. Here, we report  our study of light fragment (3 $\le$ Z $\le$ 5) emission from $\alpha$-cluster system ($^{40}$Ca$^*$) produced in $^{12}$C (77 MeV) + $^{28}$Si reaction, as well as those from the neighboring composite system  $^{39}$K$^* $ produced at the same excitation energy ($ \sim $67 MeV) via two different reaction channels  ($^{11}$B (64 MeV)+ $^{28}$Si and $^{12}$C (73 MeV)+ $^{27}$Al); the last two reactions have been chosen to crosscheck the equilibrium decay nature (absence of entrance channel dependence) of the energy damped binary fragment yield  in the decay of $^{39}$K$^* $. The time scales and the angular momentum dissipation factors for DI fragment emission in these reactions have also been studied.

The article has been arranged as follows. The experimental arrangement has been described in Sec. II. The experimental results and analysis have been presented in Sec. III and the discussions of the results have been given in Sec. IV. Finally, the summary  has been given in Sec. V.

\section{\label{sec2:expt}Experiment}

The experiment has been performed using $^{12}$C and $^{11}$B ion beams from the BARC - TIFR 14UD Pelletron accelerator at Mumbai. The $^{12}$C ion beam of energy 77 MeV   was bombarded on a self supporting $^{28}$Si target of thickness  $ \sim $1 mg/cm$^2$, to produce  $^{40}$Ca$ ^*$  at $ \sim $67 MeV of excitation energy.
In addition, the $^{12}$C ion beam of energy 73 MeV  and the $^{11}$B ion beam of energy 64 MeV were bombarded on $^{27}$Al (self supporting, $ \sim $500 $\mu$g/cm$^2$), $^{28}$Si (thickness same as above)  targets, respectively, to produce the same composite  $^{39}$K$ ^*$, at the  same excitation energy ($ \sim $67 MeV).
The fragments (3 $\le$ Z $\le$ 5) have been detected using  Silicon detector (surface barrier) telescopes ($\sim$10$\mu$m $\Delta$E, $\sim$350$\mu$m E).  The calibration of the telescopes were done using   the elastically scattered $^{12}$C,  $^{11}$B ions  from Al, Si and Au targets. The inclusive energy distributions of the emitted fragments for each reaction  have been measured in the laboratory angular range of $ \sim $12$^\circ$ to 55$^\circ$ [ $ \sim $18$^\circ$ - 82$^\circ$ in the center-of-mass (c.m.) frame]. The total systematic error in the data, arising from the uncertainties in the measurements of the solid angle, the target thickness, and the calibration of current digitizer have been estimated to be $\approx$ 12\%.

\section{\label{sec3:result}Energy spectra}

\begin{figure}
\includegraphics[scale=0.45,clip=true]{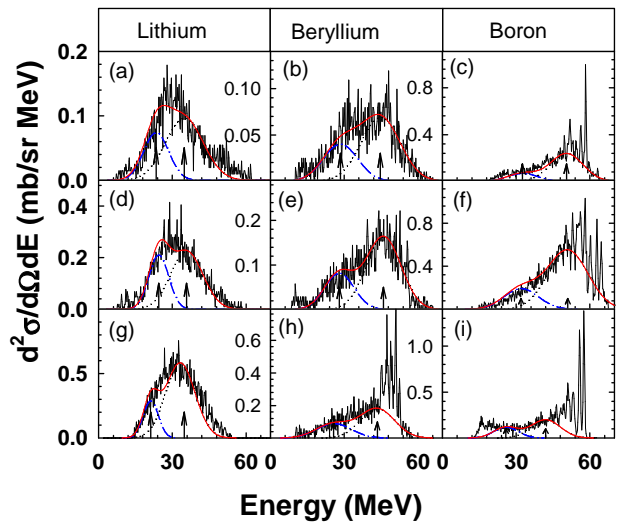}
\caption{\label{fig1:E_spec} (Color online) Typical energy spectra of the fragments measured  for the  reactions $^{12}$C  + $^{28}$Si  (a) -- (c), $^{12}$C  + $^{27}$Al (d) -- (f) and $^{11}$B  + $^{28}$ Si (g) -- (i) at  $\theta_{lab}$ = 17.5 (a) -- (h) and  30$^\circ$ (i). The blue dash-dotted, the black dotted, and the red solid curves represent the contributions of the FF, the DI, and the sum (FF + DI), respectively. The left and the right arrows correspond to the centroids of FF and DI  components of energy distributions, respectively. }
\end{figure}

Typical energy spectra of  the fragments (3 $\le$ Z $\le$ 5) emitted in  $^{11}$B (64 MeV) + $^{28}$Si, $^{12}$C (73 MeV) + $^{27}$Al and  $^{12}$C (77 MeV)+ $^{28}$Si reactions have been  shown in Fig.~\ref{fig1:E_spec}.  It is clear from the figure that the shapes of the fragment energy spectra   obtained in the  three reactions are quite different. This is mainly due to the variation of the relative contributions of DI and FF processes in each case.

In order to extract the FF and the DI components,  the energy distribution of each fragment at each angle has been fitted with two Gaussian functions in two steps, as prescribed in \cite{bhat2002,bhat2004}.  In the first step, the FF contribution has been obtained by fitting the energy distribution with a Gaussian having the centroid energy obtained from Viola systematics, duly corrected for the asymmetric factor \cite{viola,beck92}. The width of the Gaussian has been obtained by fitting the lower energy tail of the spectrum. The FF component of the energy spectrum thus obtained has then been subtracted from the full energy spectrum. In the next step, the DI component has been obtained by fitting the subtracted energy spectrum with a second Gaussian. The contributions of FF and  DI  components thus obtained (for each fragment) have been displayed  in Fig.~\ref{fig1:E_spec}.  In each spectrum, the  arrow  at lower (higher) energy  indicates   the position of the  centroid  of the FF (DI) energy distribution.

\subsection{\label{sec3:FF} Study of FF fragments }
\subsubsection{\label{sec3:AngDisFF}Angular distribution}

\begin{figure}
\includegraphics[scale=0.8,clip=true]{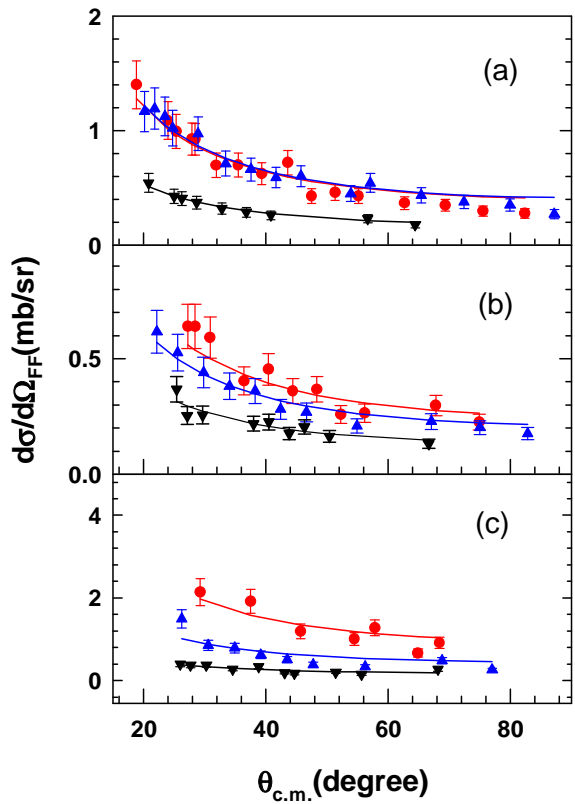}

\caption{\label{fig2:FF_AngDis} (Color online) The c.m. angular distributions  of the fragments Li (a), Be (b) and B (c). Solid circles (red), triangles (blue) and inverted triangles (black) correspond to the experimental data  for the reactions  $^{11}$B + $^{28}$Si, $^{12}$C + $^{27}$Al and  $^{12}$C + $^{28}$Si, respectively. Solid curves are fit to the data with the function  \emph{f}($\theta_{c.m.}$) $\propto$1/sin$\theta_{c.m}$.}
\end{figure}

The FF fragment angular distribution has been obtained by integrating the corresponding Gaussian extracted from the  energy distribution. The c.m. angular distributions (d$\sigma$/d$\Omega$$_{FF}$) of the FF fragments (Li, Be and B)   have been shown in Fig.~\ref{fig2:FF_AngDis}. It is evident from the figure that the  angular distributions of all FF fragments   follow $ \sim $~1/sin$\theta_{c.m.}$  dependence, which is characteristic of the fission-like decay of an equilibrated  composite system. It is also clear  from the figure that the yields of Li and Be  are  almost same at all angles for $^{11}$B + $^{28}$Si and  $^{12}$C~+~$^{27}$Al  reactions.  It has further been observed that  yield of the fragment B  in $^{11}$B~+~$^{28}$Si reaction was more than that in  $^{12}$C~+~$^{27}$Al  reaction. It has also been observed that the  fragment angular yields for the reactions $^{11}$B~+~$^{28}$Si and $^{12}$C~+~$^{27}$Al are a little higher (though nearly comparable in magnitude) than those obtained in $^{12}$C~+~$^{28}$Si reaction at the same excitation energy.

\subsubsection{\label{sec3:FF_yield}Total fragment yield}

\begin{figure}
\includegraphics[scale=0.45,clip=true]{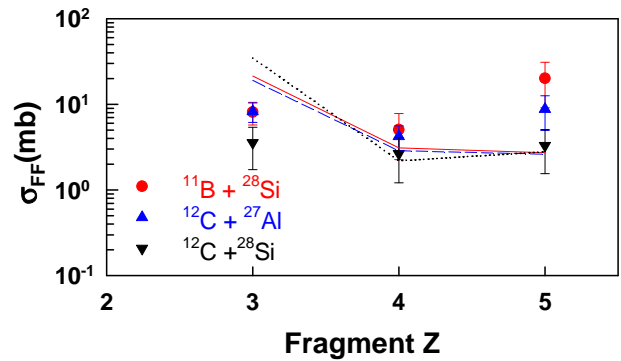}

\caption{\label{fig3:epsart} (Color online)The total FF fragment cross sections for the three reactions.  The solid circles (red), triangles (blue), and inverted triangles (black) correspond to the experimental data  for  $^{11}$B + $^{28}$Si, $^{12}$C + $^{27}$Al, and  $^{12}$C + $^{28}$Si reactions, respectively. The solid (red), dashed (blue) and dotted (black) lines are the corresponding theoretical predictions. }
\end{figure}

The experimental angle integrated  yields of the FF fragments for  all the three reactions  have been shown in Fig.~\ref{fig3:epsart}. It is found that the  yields of   Li and Be in $^{11}$B + $^{28}$Si and $^{12}$C + $^{27}$Al reactions  are nearly the same; the absence of any entrance channel dependence confirms their compound nuclear origin. It has also been observed that the  yields of these fragments  are comparable  to those obtained  in $^{12}$C + $^{28}$Si  reaction. The yield of B  in the reaction  $^{11}$B + $^{28}$Si  has been found to be slightly more than that obtained in the other two reactions, which might be due to the contamination from the beam-like channels in the former case, where B was the projectile.

The experimental FF  fragment  yields have been compared with the theoretical estimates of the same obtained from the extended Hauser-Feshbach model (EHFM) \cite{Matsuse97}. The values of the critical angular momenta have been obtained from the experimental fusion cross section data, wherever available \cite{vinyard,rousseau02}; otherwise, they have been obtained from the dynamical trajectory model calculations with realistic nucleus-nucleus interaction and the dissipative forces generated self-consistently through stochastic nucleon exchanges \cite{saila}. The values of the critical angular momentum, $l_{cr}$,  for all the three systems, have been the same  (27$\hbar$). The calculated fragment emission cross sections have been shown in Fig.~\ref{fig3:epsart}. It is seen from the figure that  in all three cases, the theoretical predictions are nearly the same and are in fair agreement with the experimental results.

\subsection{\label{sec3:DI} Study of  DI fragments}

\subsubsection{\label{sec3:DI_AngDis} Angular distribution}

\begin{figure}
\includegraphics[scale=0.38,clip=true]{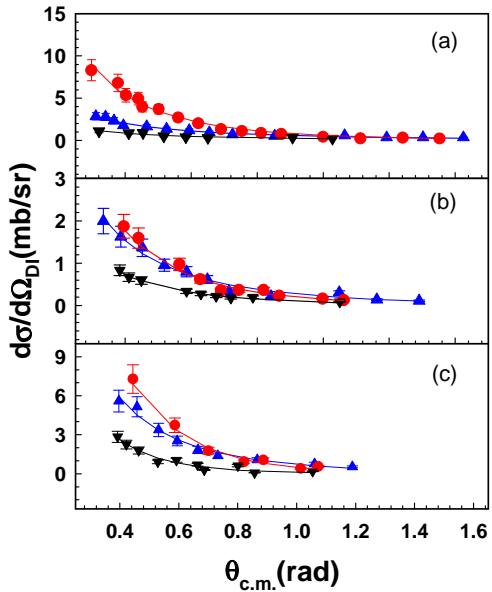}
\caption{\label{fig4:epsart}(Color online) The c.m. angular distributions of the DI  fragments [Li (a), Be (b), and, B (c)]. The solid circles (red), triangles (blue), and inverted triangles (black) correspond to the experimental data for  $^{11}$B + $^{28}$Si , $^{12}$C + $^{27}$Al, and  $^{12}$C + $^{28}$Si reactions,  respectively; the solid lines are the fits to the data (see text). }
\end{figure}

 The  DI  component of the  fragment angular  distribution   has  been  obtained  by  integrating  the  respective Gaussian extracted from the energy distribution data.  The  c.m.  angular  distributions  of  the DI  components d$\sigma$/d$\Omega$$_{DI}$  of the fragments have been  displayed in Fig.~\ref{fig4:epsart}.  It is found that they fall much faster than  $ \sim $1/sin$\theta_{c.m.}$  distribution,  indicating shorter lifetime of the composite system. Such lifetimes are incompatible with the formation of an equilibrated compound nucleus, but may  still  reflect  significant  energy  damping within the deep-inelastic collision mechanism.  It  is possible  to estimate the lifetime of the intermediate di-nuclear complex using a diffractive Regge-pole model \cite{Mikumo80, bec98} from these measured forward peaked angular  distributions. The  angular distributions have been fitted using the following expression,

\begin{equation}
               (d\sigma/d\Omega)_{DI}   =        (
C/sin\theta_{c.m.})(e^{-\theta_{c.m.}/\omega \tau_{DI}}).
\label{eq:dsig}
\end{equation}

\begin{figure}
\includegraphics[scale=.5,clip=true]{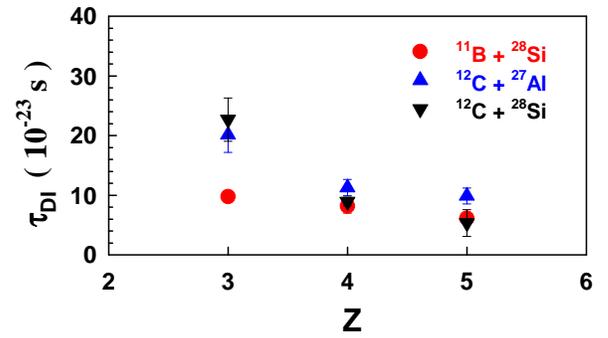}
\caption{\label{fig5:epsart} (Color online)
The emission time scales of different DI fragments. }
\end{figure}

\noindent
The expression describes  the  decay  of  a di-nucleus rotating with an angular velocity $\omega$ = $\hbar l/\mu R^{2}$, where $\mu$ is the reduced mass of the system,  $l$ is the   angular  momentum
($l_{cr} < l < l_{gr}$; $l_{gr}$, $l_{cr}$ being the  grazing and the critical  angular momenta, respectively), $R$ represents the  distance  between  the  two centres  of  the  di-nucleus and $\tau_{DI}$ is the time interval during which the two nuclei remain in a  solid  contact  in  the  form  of  the  rotating di-nucleus.   The value of the `life angle' $\alpha (= \omega \tau_{DI})$ decides the time scale of the reaction. The forward   peaked angular distributions (and small values of $\alpha$) are   associated  with  the fast  processes; on the contrary, large  values  of  $\alpha$ ( $ \gtrsim 2\pi $,  associated  with longer times as compared to the  di-nucleus  rotation  period  $\tau_{eq}$ = 2$\pi$/$\omega$), correspond to the long lived configurations and lead to  isotropic  angular distributions. The  time  scales for different DI fragments (Li, Be and B)  thus obtained have been shown in Fig.~\ref{fig5:epsart} for  comparison. It is seen that, in all reactions, the  time  scale  decreases  as  the  fragment  charge increases, which is in conformity with a previous study by Mikumo \textit{et al.} \cite{Mikumo80}. This is expected because the heavier fragments (nearer  to  the  projectile) require  less  nucleon exchange and therefore less time; on the other hand, the emission of lighter fragments requires  more nucleon exchange and therefore longer times. The emission time scales of the fragments are related to the number of nucleons exchanged on the average. This explains why the emission time scales of  $^{12}$C+$^{27}$Al and $^{12}$C+$^{28}$Si reactions are nearly the same for all fragments. On the other hand, in the case of $^{11}$B+$^{28}$Si reaction,  net nucleon exchange is one less to reach any particular fragment; so the corresponding time scales are less. For example, in terms of net nucleon exchange, the emission time scale of Li (Be) from $^{11}$B+$^{28}$Si should be comparable to that of Be (B) from $^{12}$C+$^{27}$Al and $^{12}$C+$^{28}$Si reactions, which is actually the case (Fig.~\ref{fig5:epsart}).
\vspace{.1cm}
\noindent
\subsubsection{\label{sec3:DI_Q} Average Q value}
\begin{figure}
\includegraphics[scale=.6,,clip=true]{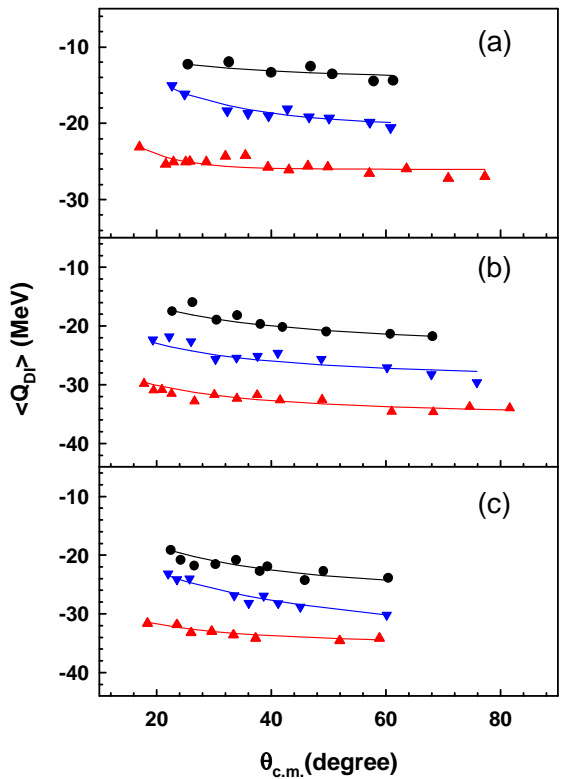}

\caption{\label{fig6:epsart} (Color online)
The  average $Q$ values, $<Q_{DI}>$, plotted as function of  $\theta_{c.m.}$ for Li (red triangle), Be (blue inverted triangle), and B (black solid circle) emitted in (a) $^{11}$B + $^{28}$Si, (b) $^{12}$C + $^{27}$Al, and (c) $^{12}$C + $^{28}$Si reactions. Solid lines are plotted to guide the eye.}
\end{figure}

The average  $Q$ values ($<Q_{DI}>$) of the  DI   fragments, estimated from the fragment kinetic energies assuming two-body
kinematics, have been displayed in Fig.~\ref{fig6:epsart}  as a function of the c.m. angle.   It is found that, for all fragments, the $<Q_{DI}>$ values  tend to decrease with the increase of angles for $\theta_{c.m.} \lesssim 40^\circ$, and then gradually become nearly constant.
It implies that, beyond this point, the kinetic energy damping is complete and dynamic equilibrium has been established before the scission of the di-nuclear composite takes place.

\subsubsection{\label{sec3:DI_yield} Total fragment yield}

\begin{figure}
\includegraphics[scale=0.45,clip=true]{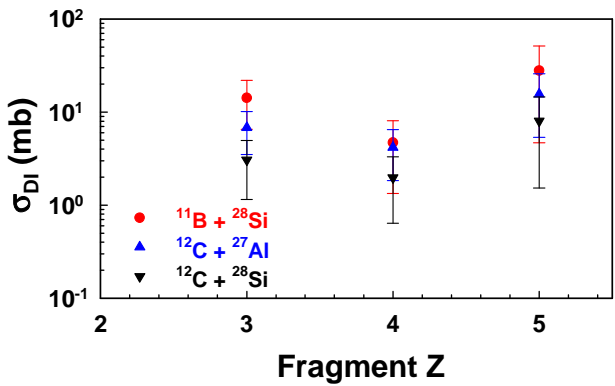}
\caption{\label{fig7:epsart}(Color online) Total DI cross sections of the fragments obtained in three different reactions.}
\end{figure}

 The experimental angle integrated  yields of the DI fragments  for $^{11}$B + $^{28}$Si, $^{12}$C  + $^{27}$Al, and,  $^{12}$C  + $^{28}$Si reactions  are shown in Fig.~\ref{fig7:epsart}.   It is found that the DI yields of all the fragments
 emitted in  B+Si reaction  are slightly higher  than  those obtained in  C+Al  and C+Si reactions. This may be due to the variation of the probability of net nucleon exchange. In addition, the DI fragment yield in C+Si reaction  tends to be lower than that for C+Al reaction.
\section{\label{sec4:discussion}Discussions}
\subsection{\label{sec:level}FF fragment emission}
 In the case of the decay of $^{40}$Ca$^*$, the measured FF fragment yields (3 $\le Z \le$ 5) have been found to be in good agreement with the respective statistical model predictions (see Fig.~\ref{fig3:epsart}), indicative of the compound nuclear origin of these fragments. However, a previous study on the binary decay of the same system  \cite{shapi82} (using inverse kinematical reaction) had reported an enhancement of fragment (6 $\le Z \le$ 8) yield over the statistical model prediction and thereby conjectured the presence of orbiting mechanism. In the case of the decay of $^{39}$K$^*$, the absence of any entrance channel dependence (between B+Si and C+Al systems) and the matching of the extracted FF fragment yields  with the  respective EHFM predictions (see Fig.~\ref{fig3:epsart}) have been clearly suggestive of  the compound nuclear origin of these fragments.

\subsection{\label{sec4:DI_f} Angular momentum dissipation factor}
\begin{figure}
\includegraphics[scale=.5]{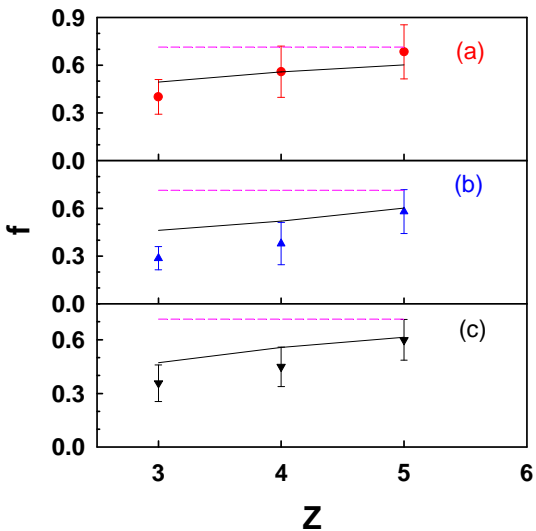}


\caption{\label{fig8:epsart}
(Color online) The variation of angular momentum dissipation factor $f$ with
fragment. The solid circles (red), solid  triangles (blue), and inverted  triangles  (black)  are the extracted values of  $f$ for (a) $^{11}$B + $^{28}$Si, (b) $^{12}$C + $^{27}$Al, and (c) $^{12}$C + $^{28}$Si reactions,  respectively. The solid (black) and
dotted (pink) curves correspond to the sticking limit and the rolling limit predictions for the same, respectively. }
\end{figure}

The angular momentum dissipation in DI collision is important to understand the variation of the mean kinetic energies of the fragments as well as the energy damping mechanism in general. For heavy systems, the angular momentum dissipation is experimentally estimated using the $\alpha$-particle angular distribution and the $\gamma$-ray multiplicity data and it is known that the rigid rotation limit is usually reached in these systems \cite{sob3}. For the light systems, the angular momentum transfer is generally  estimated from the total kinetic energy of the rotating di-nuclear system, $E_k$, which is given by,

\begin{equation}
E_k = V_N(d) + f^2 {\hbar^2 l_i (l_i+1) \over 2 \mu d^2} ,
\label{eq:ef}
\end{equation}

\noindent
where $V_N(d)$ is the contribution from Coulomb and nuclear forces at di-nuclear
separation distance $d$, $\mu$ is the reduced mass of the di-nuclear configuration,
$l_i$ is the relative angular momentum in the entrance channel and $f$ is the numerical factor denoting the fraction of the angular momentum transferred.
For these light systems, there have been indications of large dissipation of relative angular momentum \cite{shapi}, which might be partly due to the
ambiguity in the determination of the magnitude of angular
momentum dissipation, as both $d$ and $f$ are unknown quantities (see Ref. \cite{bhat2004} and references therein).
A simple prescription for  estimating both $f$ and $d$ was described in Ref. \cite{bhat2004}, where it has been shown that the fraction of angular momentum transfer for fully energy-damped DI collision of a few light systems is close to the corresponding rigid rotation limit (sticking limit). To see whether this trend is valid in general for DI collisions of light systems, angular momentum dissipation factor, $f$, for each exit channel mass asymmetry has been extracted for all the reactions, which have been displayed in Fig.~\ref{fig8:epsart}. For the present calculations, the separation distance $d$ between the two fragments
has been estimated from  the scission point configuration corresponding to the respective asymmetric mass splitting \cite{beck92}, and the value of
initial angular momentum $l_i$ has been taken to be equal to the critical
angular momentum for fusion, $l_{cr}$.

\noindent

It is observed from Fig.~\ref{fig8:epsart} that for all the three  reactions considered, the experimental values of the mean angular momentum dissipation  are more than those predicted under the
rolling condition; however, the corresponding
sticking limit predictions of $f$  are in fair agreement with  the experimental values of  the same within the error bar.  In all cases, the discrepancy is more for the lighter
fragments, and it gradually decreases for the heavier fragments. This may be explained in terms of the following qualitative argument. Microscopically, friction is generated
due to stochastic exchange of nucleons between the reacting
partners through the window formed by the overlap of
the density distributions of the two. Stronger friction  essentially means larger degree of density overlap
and more nucleon exchange. The lighter DI fragment
(corresponds to more net nucleon transfer) originates
from deeper collision, for which the interaction time is also larger as seen in Fig.~\ref{fig5:epsart}.
Therefore, the angular momentum dissipation, originating
due to the stochastic nucleon exchange, should also be more, which, at
least qualitatively, explains the observed trend. \\

\section{\label{sec5:summary}Conclusion }
 Light fragment (3 $\le Z \le$ 5)  emission in  $^{11}$B (64 MeV) + $^{28}$Si, $^{12}$C (73 MeV) + $^{27}$Al and  $^{12}$C (77 MeV) + $^{28}$Si reactions have been studied in details.  The inclusive double differential cross sections for the fragments emitted in these reactions   have been measured  in the angular range of $ \sim $12$^\circ$ to 55$^\circ$.
The energy distributions of the fragments have been fitted with two Gaussians to extract the fusion-fission and the deep-inelastic components.   The c.m. angular distributions  of the  fusion-fission fragments  have been found to follow 1/sin$\theta_{c.m.}$  dependence, which signifies the emission of these fragments  from a long-lived equilibrated composite. The total elemental cross-sections of the FF fragments have been obtained by integrating the angular distributions of the FF components. In the case of  $^{12}$C + $^{28}$Si reaction, the integrated yields of the light fragments (3 $\le Z \le$ 5) have been found to be in fair agreement with the statistical model predictions. It is interesting to note here that a previous study on fragment decay from the same system ($^{40}$Ca$^*$, produced through  inverse kinematical reaction $^{28}$Si + $^{12}$C at same excitation energy \cite{shapi82}) has shown signatures of enhancement in fragment yield (for relatively heavier fragments; 6 $\le Z \le$ 8) over those predicted by the statistical model.

We have also studied fragment emission from the  nearest non - $\alpha$ cluster system,   $^{39}$K$^* $,  produced at the same excitation energy (67 MeV) via two different entrance channels  viz.  $^{11}$B (64 MeV)+ $^{28}$Si and $^{12}$C (73 MeV)+ $^{27}$Al respectively.   It has been found that the angular distributions of the FF fragments (3 $\le Z \le$ 5) obtained in these reactions are almost similar and follow the 1/sin$\theta_{c.m.}$ dependence, indicating the emission from an equilibrated source.  The absence of any entrance channel dependence is consistent with the compound nuclear origin of these fragments.

The DI component of the fragment (3 $\le Z \le$ 5) energy distribution  in  all the three reactions has been studied in details.  It has been shown that the DI  fragment angular distribution falls much faster than   1/sin$\theta_{c.m.}$  distribution. The time scale of the DI process has been estimated from these  DI angular distributions.  It has been observed that  for all these reactions, the  time  scale, which is related to net nucleon transfer,  decreases  as  the  fragment  charge increases (closer to the projectile charge). It has also been observed that the average
$Q$ values for the DI  fragments  decrease with the increase of emission angle and saturate at higher angles, signifying a saturation in energy damping process beyond these angles. Assuming a compact exit channel configuration (estimated
from the extracted FF part of the spectra), the angular momentum
dissipation factor, $f$, for the DI process has been extracted.   For all the three  reactions,   the experimental values of $f$ have been found to be in fair agreement with the corresponding sticking limit predictions.

\begin{acknowledgments}
The authors thank the staff of the  Bhabha Atomic Research Centre, Tata Institute of
Fundamental Research (BARC - TIFR) Pelletron accelerator,  Mumbai, for smooth running of the machine.
\end{acknowledgments}

\bibliography{samir}

\end{document}